\documentclass[authoryear]{article}
\usepackage[utf8]{inputenc}
\usepackage{amsmath, amssymb, booktabs, caption, float, graphicx, hyperref, lipsum, longtable, moreverb, tikz, titling, threeparttable}

\usepackage[a4paper, total={6in, 8in}]{geometry}

\hypersetup{
 colorlinks = true, 
 urlcolor  = blue, 
 linkcolor = blue, 
 citecolor = red  
}

\begin{document}

\title{Augmenting treatment arms with external data through propensity-score weighted power-priors: an application in expanded access}

\author{ Tobias B. Polak\thanks{Department of Biostatistics, Erasmus MC, Rotterdam, The Netherlands} 
    \thanks{Department of Epidemiology, Erasmus MC, Rotterdam, The Netherlands}
    \thanks{RWD Department, myTomorrows, Amsterdam, The Netherlands}
    \thanks{School of Health Policy and Management, Erasmus University, Rotterdam, The Netherlands}
    \and Jeremy A. Labrecque\protect\footnotemark[1] \protect\footnotemark[2]
    \and  Carin A. Uyl-de Groot\protect\footnotemark[4]
    \and Joost van Rosmalen \protect\footnotemark[1] \protect\footnotemark[2]
    }
\maketitle

\abstract{The incorporation of ‘real-world data’ to supplement the analysis of trials and improve decision-making has spurred the development of statistical techniques to account for introduced confounding. Recently, ‘hybrid’ methods have been developed through which measured confounding is first attenuated via propensity scores and unmeasured confounding is addressed through (Bayesian) dynamic borrowing. Most efforts to date have focused on augmenting control arms with historical controls. Here we consider augmenting treatment arms through ‘expanded access’, which is a pathway of non-trial access to investigational medicine for patients with seriously debilitating or life-threatening illnesses. Motivated by a case study on expanded access, we developed a novel method (the ProPP) that provides a conceptually simple and easy-to-use combination of propensity score weighting and the modified power prior. Our weighting scheme is based on the estimation of the average treatment effect of the patients in the trial, with the constraint that external patients cannot receive higher weights than trial patients. The causal implications of the weighting scheme and propensity-score integrated approaches in general are discussed. In a simulation study our method compares favorably with existing (hybrid) borrowing methods in terms of precision and type-I error rate. We illustrate our method by jointly analysing individual patient data from the trial and expanded access program for vemurafenib to treat metastatic melanoma. Our method provides a double safeguard against prior-data conflict and forms a straightforward addition to evidence synthesis methods of trial and real-world (expanded access) data.}

\clearpage
\section{Introduction}
There is an increasing regulatory interest in synthesizing evidence from current (randomized) clinical trials with other data sources, to better understand the safety and efficacy of new drugs and medical devices.\cite{Reagan-UdallFoundation2018, Arlett2022} Relevant data sources include historical control arms,\cite{VanRosmalen2018,Viele2014} natural history studies,\cite{FDA2019} single-arm trials,\cite{Wang2019b} and other sources of non-trial data, such as expanded access or compassionate use programs.\cite{Polak2020, Verde2015} Ideally the incorporation of non-trial data increases power, reduces sample size, and helps to generalize results that are obtained in trial populations to more ‘real-world’ populations.\cite{Viele2014} However, the combination of trial and external data introduces several sources of potential bias that need to be attenuated via modeling strategies.\cite{VanRosmalen2018, Lewis2019} 

The variation in trial and external data can in general be attributed to either measured imbalances (e.g. in patient characteristics) between data sources and imbalances due to unmeasured confounding and other factors (e.g. center effects). Imbalances in measured characteristics can be addressed by a variety of methods such as covariate adjustment or propensity score methodology (e.g. stratification, matching or weighting). Propensity scores are frequently used to address biases that arise due to confounding in non-randomized experimental settings, by modeling allocation to treatment or control based on a set of covariates.\cite{Rosenbaum1983} However, propensity scores may also be used to distinguish between trial and external data and can thus provide a solution to the issue of confounding in the synthesis of clinical trial data and real-world data.\cite{Chen2015, Gamalo-Siebers2017a, Lin2018} 

To address unmeasured confounding, statistical methods such as (hierarchical) meta-analytical models, \cite{Neuenschwander2009, Neuenschwander2010, Schmidli2014} and the use of power-priors,\cite{Chen2000, Duan2005, Ibrahim2000} have been developed, both in frequentist and Bayesian settings. These methods perform ‘dynamic borrowing’, aiming to synthesize more evidence when data sources are ‘comparable’ and to synthesize less (or completely exclude evidence) as data sources differ increasingly. These synthesis methods were primarily developed to combine randomized controls with historical controls. In that context, Pocock suggested strict conditions relating to study design and patient characteristics to ensure that the historical data and the current trial are sufficiently comparable prior to performing a combined analysis.\cite{Pocock1976} One of Pocock's criteria is that the patient characteristics of the historical and randomized controls have a similar distributionzsk, which may not be realistic in the context of non-trial, real-world data.\cite{Hatswell2020}

Ample recent scholarship has been devoted to developing methods that simultaneously address both sources of bias. In these 'hybrid' approaches, propensity score methods are integrated into dynamic borrowing methods.\cite{Wang2019b, Lin2019} Multiply the number of standard propensity score methods (e.g. stratification, matching, weighting) with the number of available borrowing methods (such as the modified power prior, the meta-analytic predictive prior and the commensurate prior), and one may quickly get lost in the statistical jungle. In this paper, we aim to combine both fields of research in an understandable manner, and we develop a conceptually simple and easy-to-use combination of the modified power-prior with propensity score weighting. In addition we give a detailed interpretation of the entirety of 'hybrid' approaches in the framework of causal estimands. Finally, we evaluate our methods through simulation and a case study by jointly analysing the trial and expanded access program of vemurafenib in the treatment of metastatic melanoma. 

The majority of aforementioned applications focus on the integration of external (historical) controls with current trial controls. Limited attention has been devoted to research on augmenting current treatment arms with external treatment arms. This lack of research may in part be attributed to the focus on trial design for regulatory product approval. After all, it may be difficult to find an external data set on active treatment usage before the product is readily on the market. Nonetheless, these data may be available through expanded access programs. In expanded access (also known as compassionate use or early access), patients who are ineligible for registered treatment options and ongoing trials may be granted access to active, unlicensed treatments prior to regulatory approval. Expanded access pathways have become increasingly popular in recent years, and data generated through expanded access form a substantial and increasing area of academic literature - especially due to the COVID-19 pandemic. Moreover, the analyses of such access programs have been integrated into regulatory and cost-effectiveness decision making.\cite{Polak2020, Polak2022} However, the statistical literature has not yet focused on models designed for the analysis of these types of programs and through this paper, we aim to make a first contribution to this area.

The remainder of this paper is organized as follows. Section 2 discusses the background of propensity scores, dynamic borrowing, and hybrid methods. Section 3 details out our new proposed method. Section 4 evaluates our method with a simulation study, and Section 5 illustrates our method with a real-life expanded access program and trial. Finally, Section 6 concludes with a discussion. 
\newpage
\section{Background of methodology}
\subsection{Notation}
The data consist of a current (internal) trial $\mathcal{Y}_0$, and data from an external source $\mathcal{Y}_e$. In total, we have data on $N = N_0 + N_e$ patients. For every patient $i, i=1,\ldots, N$, in either the current study or the external source, we observe the outcome $y_i$, a realization of $Y_i$, and the covariate vector $x_i$ of length $K$, which is a realization of the set of covariates $\mathcal{X}$. Let $Z$ be an indicator variable, where $z_i=1$ if patient $i$ belongs to the internal study and $z_i=0$ if patient $i$ belongs to the external data source. In our case study, the estimand is the baseline rate in a single-arm study and hence there is no treatment effect. 

\subsection{Propensity scores}
Propensity scores are frequently used to address biases that arise due to confounding in non-randomized experimental settings,\cite{Rosenbaum1983} by modeling the allocation to treatment $(T=1)$ or control $(T=0)$ as a function of the covariates that one wishes to balance across these two groups:
\begin{align} \label{eq: propensity def}
   e(x) = \text{Pr}(T = 1 \mid \mathcal{X} = x).
\end{align}
Among patients with the same propensity scores $e(x)$, covariates included in the the propensity score will be balanced across the treated and untreated groups. Under the assumption that the variables in $\mathcal{X}$ are sufficient to make the treatment groups conditionally exchangeable ($Y^t \perp\!\!\!\!\perp T|\mathcal{X}$), the propensity score can be used to estimate the causal effect of treatment. Weighting, matching, and stratification are the main methods in the propensity score toolbox. \cite{Austin2011}

To use the propensity score to compare current trial data with external data, several authors have slightly redefined the propensity score.\cite{Lin2018, Lin2019, Wang2019b} Instead of modeling assignment to a control or treatment group, the propensity score is now used to model the allocation between current and external data $(Z)$:
\begin{align} \label{eq: propensity powerp def}
  \lambda_i = \text{Pr}(Z = 1 \mid \mathcal{X} = x_i),
\end{align}
where $\lambda_i$ is the probability of patient $i$ being in the internal study given the patient characteristics. Now, patients with similar propensity scores are equally likely to have been in the trial or external data conditional on $X$. If the variables in $X$ are sufficient to satisfy $Y^{t=0} \perp\!\!\!\!\perp Z|T=0,X=x$, then the internal and external populations are exchangeable.

\subsection{The power prior}
The power prior is one of the most prominent methods for dynamically borrowing information from the external data to aid inference of the current trial. The amount of borrowing - and hence the dynamic aspect - is based on how comparable the external data are to the current data. The more alike they are, the more is borrowed. An excellent review of these methods is provided by Viele and others.\cite{Viele2014} The power prior is a Bayesian methodology that incorporates the external data into an informative prior to facilitate the analysis of the current study. In this informative prior the external data is downweighted by raising its likelihood to a power parameter $\delta$, where the value of $\delta$ (with $0\leq \delta \leq 1$) controls the amount of borrowing:
\begin{align} \label{eq: powerprior}
  p(\theta \mid \mathcal{Y}, \delta) \propto \mathcal{L}(\theta | \mathcal{Y}_0) \mathcal{L}(\theta | \mathcal{Y}_e )^{\delta} \pi(\theta).
\end{align}
In the above specification, $\delta = 1$ results in a simple pooling of the two data sources, whereas $\delta=0$ effectively ignores the external data.
As it is unclear how $\delta$ should be chosen, Duan together with Ibrahim and Chen have proposed to estimate this in a fully Bayesian way,\cite{Ibrahim2000, Duan2005} in the so-called 'modified power prior' (MPP). This leads to: 
\begin{align}\label{eq: modified powerprior1}
p(\theta, \delta \mid \mathcal{Y} ) \propto \mathcal{L}(\theta | \mathcal{Y}_0) \mathcal{L}(\theta | \mathcal{Y}_e )^{\delta}  \frac{1}{C(\delta)} \pi(\delta) \pi(\theta),
\end{align}
where $C(\delta) = \int_{\theta} \mathcal{L}(\theta \mid \mathcal{D}_e)^{\delta} \pi(\theta) d \theta $ is a scaling constant to ensure \ref{eq: modified powerprior1} abides by the likelihood principle. Reviews of different power-prior specifications and their characteristics can be found in Van Rosmalen et al. or Ibrahim and Chen. \cite{Ibrahim2015, VanRosmalen2018}

\newpage
\section{Methods}
\subsection{Integrating propensity scores and power prior}
Recently, various researchers have proposed ‘propensity-score integrated hybrid approaches’, which combine propensity score methodology with dynamic borrowing methods. Methods have been developed that focus on combining propensity score stratification with power priors,\cite{Wang2019a, Wang2019b} or meta-analytic predictive priors.\cite{Liu2021} Other methods focus on the inclusion of propensity score-matching in dynamic borrowing.\cite{Lin2018} Finally, a recent review of several of these methods has proposed both propensity score-weighting together with fixed and commensurate priors.\cite{Wang2022} All these methods focus primarily on augmented control designs, designs in which the control arm of a trial is combined with external data on (historical) control arms. 

The main rationale for all these methods is the dual safeguard mechanism within the two-stage analysis: observed confounding is addressed by using propensity score methods in the first stage, and remaining unobserved confounding is attenuated via dynamic borrowing methodology in the second stage. 

We add to this literature by proposing a novel method based on propensity score weighting and the modified power-prior to augment the current treatment arm with external treatment data. The basis of our method, which we refer to as the ProPP, is the modified power prior, which is designed to only address imbalances due to unmeasured confounding (see \ref{eq: modified powerprior1}). To also safeguard against the effects of measured imbalances in patient characteristics, we apply propensity score weighting to this likelihood function before it is used in the MPP. The propensity-score weighted likelihood function is given by
\begin{align} \label{eq: weightedlikelihoodgen}
  \mathcal{L}(\theta \mid \mathcal{Y}) &=\prod_i f(y_i \mid \theta)^{w_i},
\end{align}
where $w_i$, the weight used for patient $i$, is chosen as a function of the propensity score $\lambda_i$. If we now substitute this likelihood for the external data in \ref{eq: weightedlikelihoodgen}, we obtain:
\begin{align} \label{eq: weightedlikelihoodext}
  \mathcal{L}(\theta \mid \mathcal{Y}_e)^{\delta} &= \left( \prod_i f(y_i \mid \theta)^{w_i} \right) ^{\delta} ,
\end{align}
Because our method combines propensity scores with dynamic borrowing based on the MPP, the effective weight for patient $i$ is $\delta \times w_i$. In our approach neither $\delta$ nor $w_i$ are allowed to take values greater than 1, so that the proposed method is always more conservative (i.e., provides additional protections against prior-data conflict) than the modified power prior. We additionally assess the causal and practical implications of the choice of weighting schemes.

\subsection{Causal interpretations}
\begin{figure}[ht!]
\centering
\begin{tikzpicture}
\node[text centered] at (0,0) (x) {$T$};
\node[text centered] at (0,4) (c2) {$C_1$};
\node[text centered] at (5,4) (c3) {$C_2$};
\node[text centered] at (-2,2) (c4) {$C_3$};
\node[text centered] at (0,2) (z) {$Z$};
\node[text centered] at (5,0) (y) {$Y$};

\draw[->, line width= 1] (x) -- (y);
\draw[->, line width= 1] (z) -- (x);
\draw[->, line width= 1] (c3) -- (y);
\draw[->, line width= 1] (c2) -- (y);
\draw[->, line width= 1] (c2) -- (z);
\draw[->, line width= 1] (c4) -- (z);
\end{tikzpicture} 
\caption{Directed acyclic graph to explore the causal implication of combining propensity scores and dynamic borrowing methods}
\label{figure:DAG}
\end{figure}
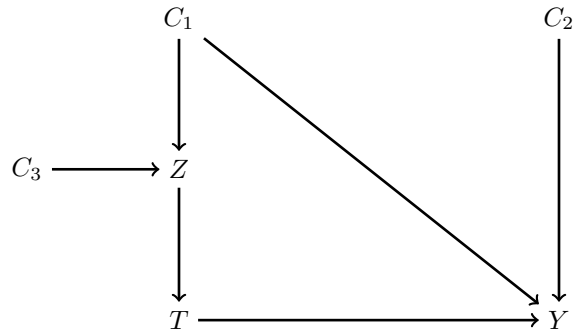

 The weights $w_i$ can be chosen in a variety of ways. In applications of propensity score weighting for the estimation of treatment effects in observational studies, the weights are typically allowed to vary between patients within each group and depend on the estimand of interest. Before we choose $w_i$, we here want to provide an explicit causal interpretation of different modeling choices, or different choices of weights $w_i$ (see Figure \ref{figure:DAG}).
 
 The value of $Y$ can be different in $Z=1$ and $Z=0$ either due to random error or due to confounding (or selection bias) between $Z$ and $Y$. An example of such a confounder would be any cause of the outcome that is not equally distributed in the internal and external data (e.g., $C_1$ in Figure 1). Dynamic borrowing methods based on differences in the outcome aim to balance the risk of pooling data with systematic differences (i.e., due to confounding) with the benefit of pooling data with differences only due to random error which increases precision. In this way, dynamic borrowing can never eliminate bias due to a variable such as $C_1$ but it can attenuate the bias by reducing the degree of pooling.
 
Ideally, differences in $Y$ due to variables such as $C_1$ would be removed before dynamic borrowing determines the degree of pooling. Doing so would improve the bias and precision of our estimate. First, it would remove the bias due to $C_1$. Second, if $C_1$ increases the differences in $Y$ across levels of $Z$, removing the effect of $C_1$ would reduce this difference and would therefore increase the degree of pooling while not sacrificing validity. The goal of the propensity score weights is precisely this: they re-weight the external data in such a way that the distribution of the variables used to compute the propensity score is the same across $Z$. In the re-weighted population there is no longer any relationship between the variables in the propensity score and $Z$. In Figure 1, weights constructed from the propensity score estimated using $C_1$ would in essence remove the edge between $C_1$ and $Z$. Of course, the propensity score-based weights only balance variables used to construct the propensity score. Any confounders which are not included in the propensity score will remain unbalanced across $Z$ even in the re-weighted data set. As was the case before considering propensity score-based weighting, dynamic borrowing can balance the risk of unmeasured confounding with the benefit of increased precision, but with the additional benefit that some of the systematic difference in $Y$ across levels of $Z$ due to confounding has been removed through the propensity score-based weighting process. 
 
 Some caution is required when considering which variables to include in the propensity score model. Variables that are a cause of the outcome but unrelated to $Z$ ($C_2$, Figure 1), are not necessary to include but they may help increase precision. Variables related to $Z$ but not directly related to $Y$ ($C_3$, Figure 1) should not be included and may in fact amplify any bias due to uncontrolled confounding between $Z$ and $Y$.

\subsubsection{The choice of weights}
In dynamic borrowing, the usual goal is to improve the estimate of the outcomes from the trial. We should choose $w_i$ according to a weighting scheme that corresponds with our estimand of interest, i.e. the average causal effect among those in the trial should be our target estimand (see Table \ref{table:weightingschemes}). Therefore, we use a weighting scheme based that targets that estimand.
\begin{table}[ht!] 
\centering \caption{Propensity score weighting schemes under different populations of interest.}
\begin{tabular}{@{}llll@{}}
\toprule
Trial        & External        & Population of interest                  & \\ \midrule
$\frac{1}{\lambda(x)}$   & $\frac{1}{1-\lambda(x)}$  & Average treatment effect        & \\
1           & $\frac{\lambda(x)}{1-\lambda(x)}$ & Average treatment effect of the trial & \\
$\frac{1-\lambda(x)}{\lambda(x)}$ & 1           & Average treatment effect of the external & \\ \bottomrule
\end{tabular}
\label{table:weightingschemes}
\end{table}
However, we slightly adapt the above weighting scheme to make sure that no subject in the external data obtains a weight larger than 1. Weights larger than 1 would be undesirable for two reasons. First, this would amount to an inflation of the sample size in a Bayesian analysis, which in turn would lead to an overestimation in the precision of the estimates. Second, weighting a non-trial participant higher than current trial participants may cause regulatory concerns. Therefore, we propose to maximize the weight of the patients in the external data set at 1 (meaning that this patient is equally likely to have come from the trial). We set the weight $w_i$ of all trial patients to 1, and of all external patients to: $w_i = \min\{1, \frac{\lambda(x)}{1-\lambda(x)} \}$. 

\subsection{Implementation for dichotomous outcomes}
Here we illustrate the implementation of our method for data with a Bernoulli-distributed dichotomous outcome measure, with mean $\theta$ and likelihood function (before applying propensity score weights) given by $L(\theta \mid y_i) = \theta^{y_i}(1-\theta)^{1-y_i}$. Filling in this expression in \eqref{eq: weightedlikelihoodgen} gives the propensity score weighted likelihood function:
\begin{align} \label{eq1: derivation}
 \mathcal{L}(\theta \mid Y, w_i) = \theta^{\delta \sum w_i y_i} (1-\theta)^{\delta \sum w_i (1-y_i)}. 
\end{align}
Combining this propensity score weighted likelihood function with the posterior of the modified power prior in \eqref{eq: modified powerprior1} gives the joint posterior of the power parameter $\delta$ and the mean $\theta$ as
 \begin{align*}
\pi(\theta, \delta \mid Y, w_i) = \theta^{\delta \sum w_i y_i^e + \sum w_i y_i^0} (1-\theta)^{\delta \sum w_i (1-y_i^e) + \sum w_i (1-y_i^0) } \frac{1}{C(\delta)} \pi(\delta) \pi(\theta).
\end{align*}
With a uniform $U(0,1)$ prior for the mean parameter $\theta$ the integral in the scaling constant can be solved analytically as
\begin{align*}
C(\delta) &= \int_{\theta} \mathcal{L}(\theta \mid \mathcal{Y}_e, w_i)^{\delta} \pi(\theta) d \theta = B \left( \delta \sum w_i y_i^e + 1, \delta \sum w_i (1-y_i^e) + 1 \right).
\end{align*}
If we further assume a $\textrm{Beta}(\alpha_\delta,\beta_\delta)$ prior for $\delta$, the joint posterior becomes
\begin{align}\label{posteriorProPP}
\pi(\theta, \delta \mid Y , w_i) = \displaystyle \frac{\theta^{\delta \sum w_i y_i^e + \sum w_i y_i^0 + \alpha_\delta} (1-\theta)^{\delta \sum w_i (1-y_i^e) + \sum w_i (1-y_i^0) + \beta_\delta }}{B \left( \delta \sum w_i y_i^e + 1, \delta \sum w_i (1-y_i^e) + 1 \right)}
\end{align}
and after integrating out $\theta$, the marginal posterior of $
\delta$ is given by 
\begin{align}\label{marginalposterior}
\pi(\delta \mid Y, w_i) &= \displaystyle \frac{B ( \delta \sum w_i y_i^e + \sum w_i y_i^0 + \alpha_\delta, \delta \sum w_i (1-y_i^e) + \sum w_i (1-y_i^0) + \beta_\delta )}{B \left( \delta \sum w_i y_i^e + 1, \delta \sum w_i (1-y_i^e) + 1 \right)}.
\end{align}
With the assumed prior distributions, the conditional posterior of $\theta$ given $\delta$ also has a closed-form expression, namely
\begin{align}\label{conditionaldistribution}
  \theta \mid \delta, Y, w_i \sim B(\delta \sum w_i y_i^e + \sum w_i y_i^0 + 1, (1-\theta)^{\delta \sum w_i (1-y_i^e) + \sum w_i (1-y_i^0)} + 1),
\end{align}
which greatly simplifies posterior sampling.

\subsection{Algorithm} 
The sampling algorithm for the ProPP can be specified as follows:
\begin{enumerate}
  \item Obtain the propensity scores as the fitted probabilities from a logistic regression for the allocation between current and external data, based on equation \eqref{eq: propensity def}.
  \item Based on the population of interest and regulatory and statistical properties, choose a suitable weighting scheme from Table \ref{table:weightingschemes} to rescale the probabilities obtained in Step 1.
  \item Draw a sample of $\delta$ from a uniform $U(0,1)$ distribution and accept the values in that sample with probability given by \eqref{marginalposterior}; other values in the sample are removed.
  \item Draw a sample of $\theta$ from the conditional distribution \eqref{conditionaldistribution}, using the accepted values of $\delta$ from Step 3.
\end{enumerate}
In Step 3, we use a sample of size 10,000, which should suffice because the rejection sampling method used in this step generates a random (independent) sample. This sampling algorithm is easy to program, and the code for the analyses in this paper can be downloaded from the GitHub of the first author. \footnote{https://GitHub.com/TobiasPolak} 

\newpage
\section{Simulation Study}
\subsection{Setup}

We implement a simulation design to investigate the performance of our proposed method. The aim of this simulation study is to evaluate our proposed method and compare it with traditional approaches. Our simulation design was inspired by previous hybrid setups,\cite{Wang2019, Wang2019a} as well as motivated by the available setting of a (single-arm) clinical trial with external data from an expanded access program.

\subsubsection{Data generation}
We simulate the dichotomous outcome through the following data generating process:
\begin{align}
  \text{logit } y_i|x_i, z_i = \beta_0 + \boldsymbol{\beta} x_i + \eta \times I(z_i=1), 
\end{align} 
where $\beta_0$ is the intercept, $\boldsymbol{\beta}$ is a row vector of coefficients and $\eta$ is a drift term. In our base case setting, we simulate data from $N=800$ patients ($N_0=400$ in the trial, $N_e=400$ in the external data), for $K = |X| = 5$ different continuous covariates $X$ with $\beta_j = 0.1$, $j=1,\ldots,5$. We set our base case intercept to $\beta_0 = 0$. 

Several scenarios are explored to take into account that differences between trial and external outcomes can occur due to differences in covariates and/or a difference in the drift parameter. For the patient characteristics in the current trial, we assume normally distributed covariates with $X_{0} \sim \mathcal{N}(\mu_0, \sigma_0^2)$. To account for possible differences in the covariate distribution in the external data, we assume that a proportion ($\psi$) of the patients in the external data have the same covariate distribution as the trial patients, and that the other external patients ($1-\psi$) have data from a different normal distribution, with $X_{e} \sim (1-\psi) \mathcal{N}(\mu_e , \sigma_e^2) + \psi \mathcal{N}(\mu_0, \sigma_0^2)$. We vary the value of $\psi$ from 0.5 to 1 in the simulations, to assess the implications of our methods when covariates have different degrees of overlap. 

To investigate the performance of our method, we consider the following four main scenarios:
\begin{enumerate}
  \item Scenario 1: Drift. The change in outcome is only caused by drift $\eta$. We vary $\eta \in [-0.5, 0.5]$. Both populations have the same covariate distribution, i.e. $X_e, X_0 \sim N(0,1)$, but these have no effect on the outcome distributions as $\beta = 0$.
  \item Scenario 2: Mixture. The change in outcome is only caused ($\beta = 0.1)$ by a difference in the underlying covariate distributions. The covariates come from a mixture distribution with $\psi = 0.5$. We assume $X_0 \sim \mathcal{N}(0, 1)$ and $X_e \sim \mathcal{N}(\mu_e, 1)$, where we vary $\mu_e \in [-0.5, 0.5]$. There is no drift, $\eta$ = 0.
\end{enumerate}
Within these two scenarios, we also assess the following four settings: 
\begin{enumerate}
  \item Setting 1: Equal sample sizes. $N_0 = N_e = 400$ 
  \item Setting 2: Larger external data. $N_0 = \frac{1}{5} N_e = 400$ 
  \item Setting 3: Larger current trial data. $N_0 = 2 \times N_e = 400$ 
  \item Setting 4: Increase in the number of covariates, with 10 instead of 5 covariates
\end{enumerate}
Additionally, we look at how sensitive our method is to (mis)-specification. Therefore, we also consider:
\begin{enumerate}
  \item Scenario 3: No Mixture. The change in outcome is only caused by a difference in underlying covariate distributions. Unlike Scenario 2, there are no latent classes.
  \item Scenario 4: Superfluous covariates. This setting mimics setting 1, but now some of the parameters $\beta_j$ are forced to zero to simulate the inclusion of 'superfluous covariates' (i.e. $C_3$ in Figure \ref{figure:DAG}).
\end{enumerate}

Our parameter of primary interest is the baseline trial rate, $\beta_0$. Both in our simulation and in our expanded access use case, this is the response rate in a single-arm trial. 

\subsubsection{Methods and performance measure}
The methods that we compare in our simulation study belong to the following three classes: 'naive methods' such as (i) Ignore: leaving out external data and (ii) Pooling: directly combining current trial and external data, 'dynamic borrowing methods' such as  (iii) the modified power prior, and 'hybrid methods' such as (iv) the stratification + power-prior method suggested by Wang et al. \cite{Wang2019b} whilst borrowing at most 10\% and 20\% (of the current trial) of patients from the external data source. Our proposed method forms an addition to the hybrid methods.
Performance will be assessed by measuring:
\begin{align}
  \text{RMSE} &= \mathbb{E}_{\beta_0} \left[ (\beta_0 - \hat{\beta_0})^2 \right]^{\frac{1}{2}}
\end{align}
and the type I error rate. To assess the type I error rate, we checked how frequently  the objective response rate from the trial (through $\beta_0=0$) was within the equal-tailed 95\% posterior credible interval of our estimand. 

\subsection{Results}
\subsubsection{Scenario 1: Drift} \label{sec: Drift}
In scenario 1 'Drift', patients from the trial are similar to patients from the external data (i.e. their covariates come from the same underlying distribution), but the outcomes differ due to a random drift term $\delta$. The scenario of drift is the standard situation where methods such as the MPP are usually evaluated. The results for type I error rate and RMSE are shown in Figure 1.B. The RMSE of the analysis without external data (Ignore) is approximately 0.034. In case there is no drift, pooling the two data sources gives the lowest RMSE (approximately 0.023), 32$\%$ lower than ignoring the external data. The RMSE of pooling increases considerably when there is a nonzero drift, e.g. with a drift of $\delta = 0.375$, the RMSE of pooling is 0.05 - a 47$\%$ increase compared with ignoring external data, and the type I error rate becomes severely inflated. 

\begin{figure}[H] 
\centerline{\includegraphics[scale=0.50]{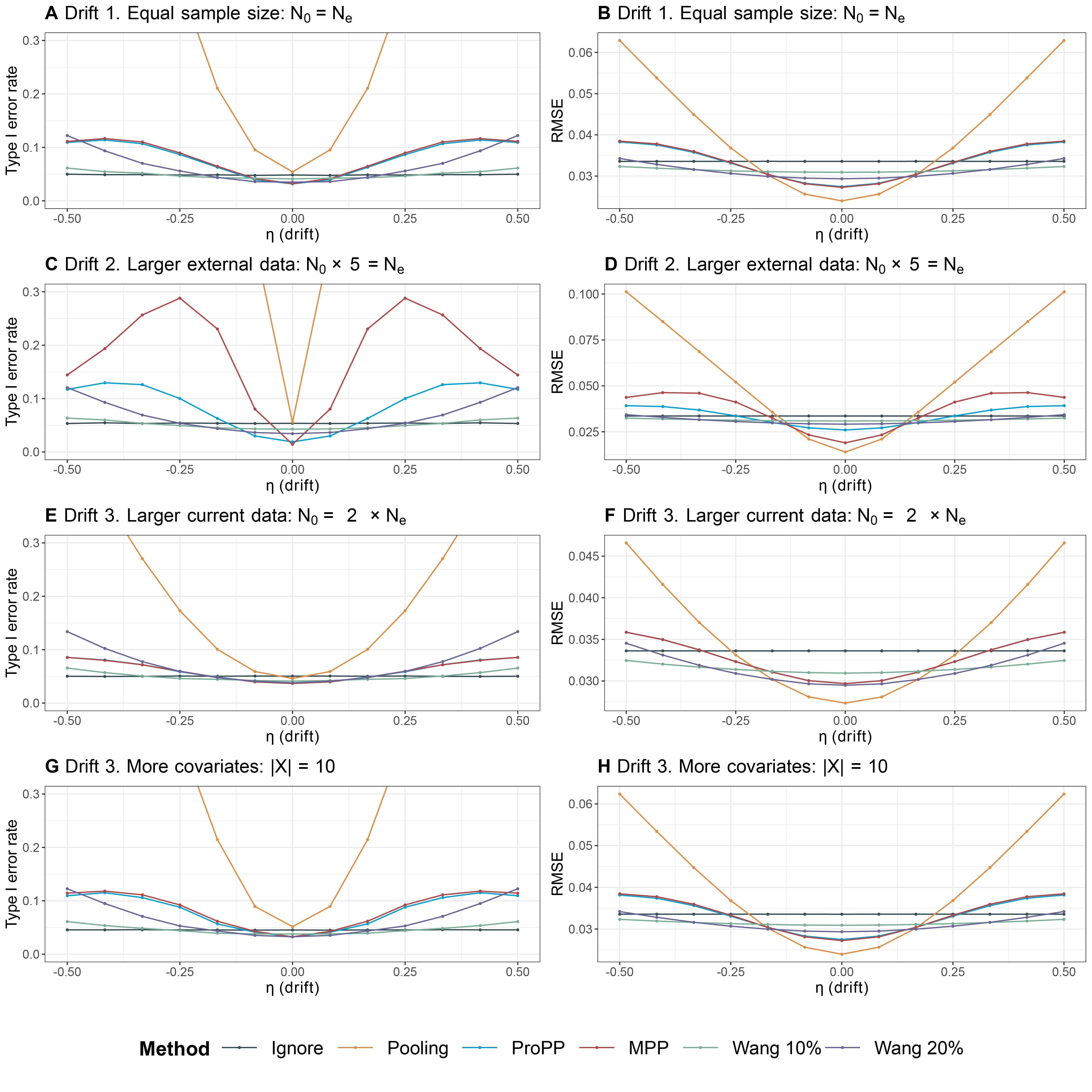}}
\centering 
\caption{Comparison of different estimation methods in terms of Type-I error rate (left) and root mean squared error (RMSE) (right) when the difference in outcomes is in part driven by a random drift term. There is no difference covariates (Setting 1).}
\label{Figure: Drift}
\end{figure}

For all cases except sub-setting 2, the RMSE and type I error rates of the ProPP and the MPP almost overlap and show the same characteristics (see Figure \ref{Figure: Drift}). In this scenario, where sample sizes are equal and patients are similar, all patients have approximately a probability of $\frac{1}{2}$ to be in the trial or the external data (and hence odds $w_i$ of $0.5/(1-0.5) = 1$). When $w_i = 1$ for all patients, the ProPP specification (\ref{eq: weightedlikelihoodext}) simplifies to the MPP specification in (\ref{eq: modified powerprior1}). Sub-setting 2 $(N_0 = 400, N_e = 2000)$ shows that a relatively larger sample size in the external data causes the 'Pooling' and the 'MPP' methods to exhibit an increased RMSE and an inflated type I error rate (up to 25 percent in the MPP). The weights $w_i$ in the ProPP naturally account for such a difference in sample size and prevent this (unwanted) behavior. Compared with the hybrid methods of Wang, our method has a lower RMSE, at the cost of an inflated type I error rate. Due to the pre-specified amount of borrowing, Wang's methods show a stricter control of the type I error rate in the simulations, but unlike the MPP and ProPP, this inflation continues to increase for higher values of drift, because the amount of borrowing is preset in these methods (see, for example, Wang 20 $\%$ in Figure 2A).

The results of this scenario show that the MPP and the ProPP have similar performance in terms of mitigating prior-data conflict due to unmeasured confounding. Note that the ProPP provides additional safeguards against measured confounding, which by design did not occur in this scenario. The fact that the amount of borrowed external data in the ProPP does not automatically increase with the sample size of the external data, in which this method differs from the MPP, seems an advantage. the hybrid methods proposed by Wang, the ProPP has lower RMSE, but this comes at the cost of a type I error rate inflation.

\subsubsection{Scenario 2: Mixture} \label{sec: Mixture}
\begin{figure}[H]
\centerline{\includegraphics[scale=0.50]{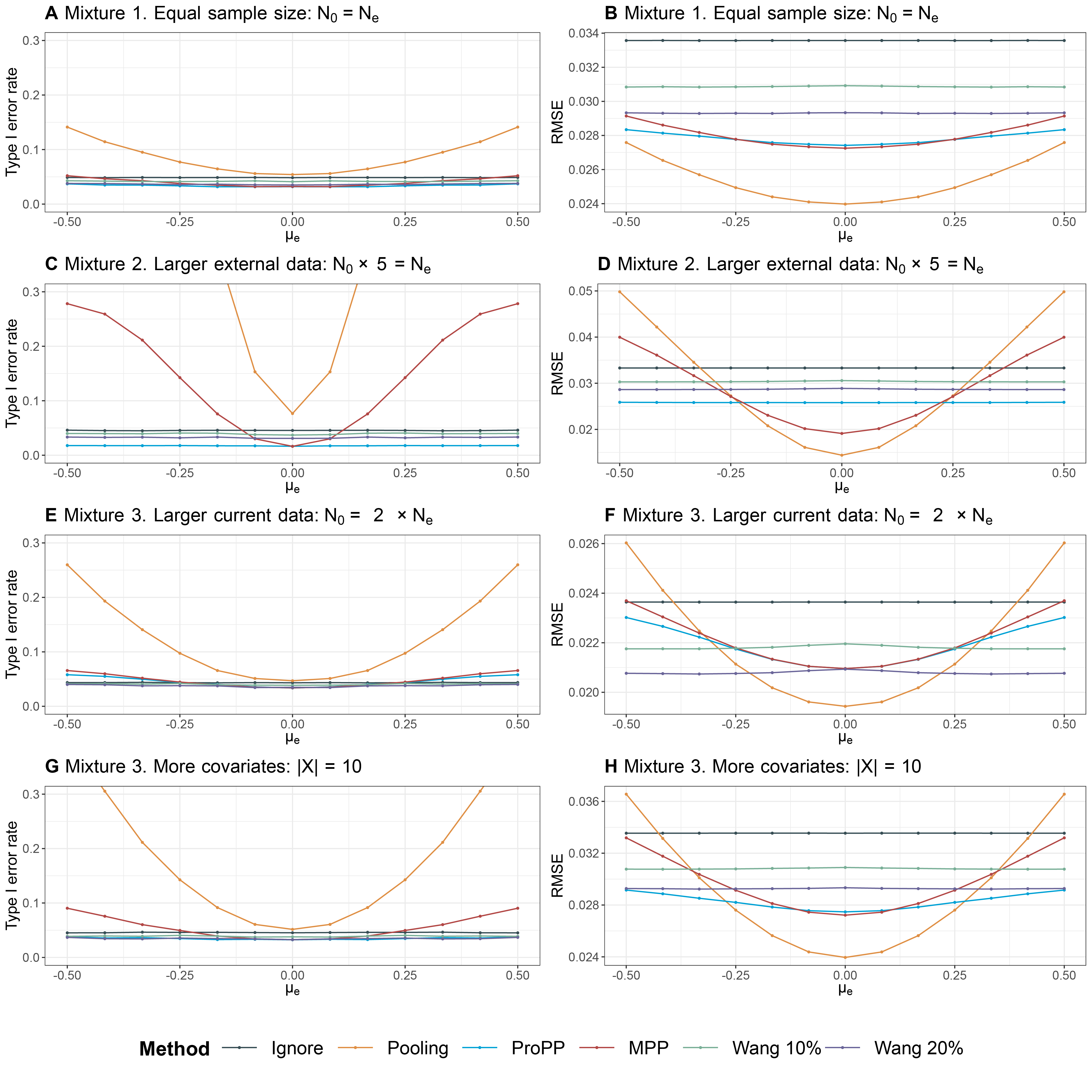}}
\centering 
\caption{Comparison of different estimation methods in terms of Type-I error rate (left) and root mean squared error (RMSE) (right) when the difference in outcomes is in part driven by difference in covariates. There is a latent class structure in covariates.(Setting 2).}
\label{Figure: Mixture}
\end{figure}

In scenario 2 'Mixture', the differences in outcomes between the current data and external data are caused by a difference in covariate distributions between patient populations. For this situation, inclusion of covariates ought to improve the operating characteristics compared with excluding covariates (like in the MPP). Figure \ref{Figure: Mixture} presents the simulation results, and the RMSEs are depicted on the right-hand side. The RMSE of Ignore is a flat line at approximately 0.034 as there is no borrowing regardless of the outcome of the external data. Both Wang 10\% (at 0.031) and Wang 20\% (at 0.029) are also relative flat lines. Pooling reaches the lowest RMSE at 0.024 in non-zero drift, followed by the MPP and the ProPP at 0.0273 and 0.0274, respectively. Both the MPP and ProPP do show an increase in type I error rate error, but remain more precise than Wang's methodology across our simulation range. By accounting for covariate effects using propensity score (i.e. rightly only incorporating similar patients), all hybrid methods yield a relatively stable and well-controlled type I error rate. This result is most clearly seen in Figure 2A, where both naive methods suffer from a large increase in type I error rate compared with the hybrid methods. 

The results of this scenario show that the incorporation of covariates through propensity score methods provides an edge over the Pooling and MPP methods. The lower RMSE of these methods compared with ignoring external data is driven by the external patients that are similar to the current patients - and exactly these similar patients receive a higher weight. By including primarily similar patients, our estimate is improved. When there are more external patients to choose from (setting 2), the chances of selecting the most similar patients increase, and the gain in precision becomes almost completely stable across settings. The increase in precision in Wang's method is driven purely by the prespecified amount of borrowing, whereas our method seems not to be impeded by borrowing limits, generally leading to a lower RMSE. Only in the unlikely setting of smaller external than current data, the ProPP relatively underperforms compared with Wangs methods - but still outperforms the MPP. 

\subsubsection{Sensitivity analysis: no mixture}
\begin{figure}[H]
\centerline{\includegraphics[scale=0.4]{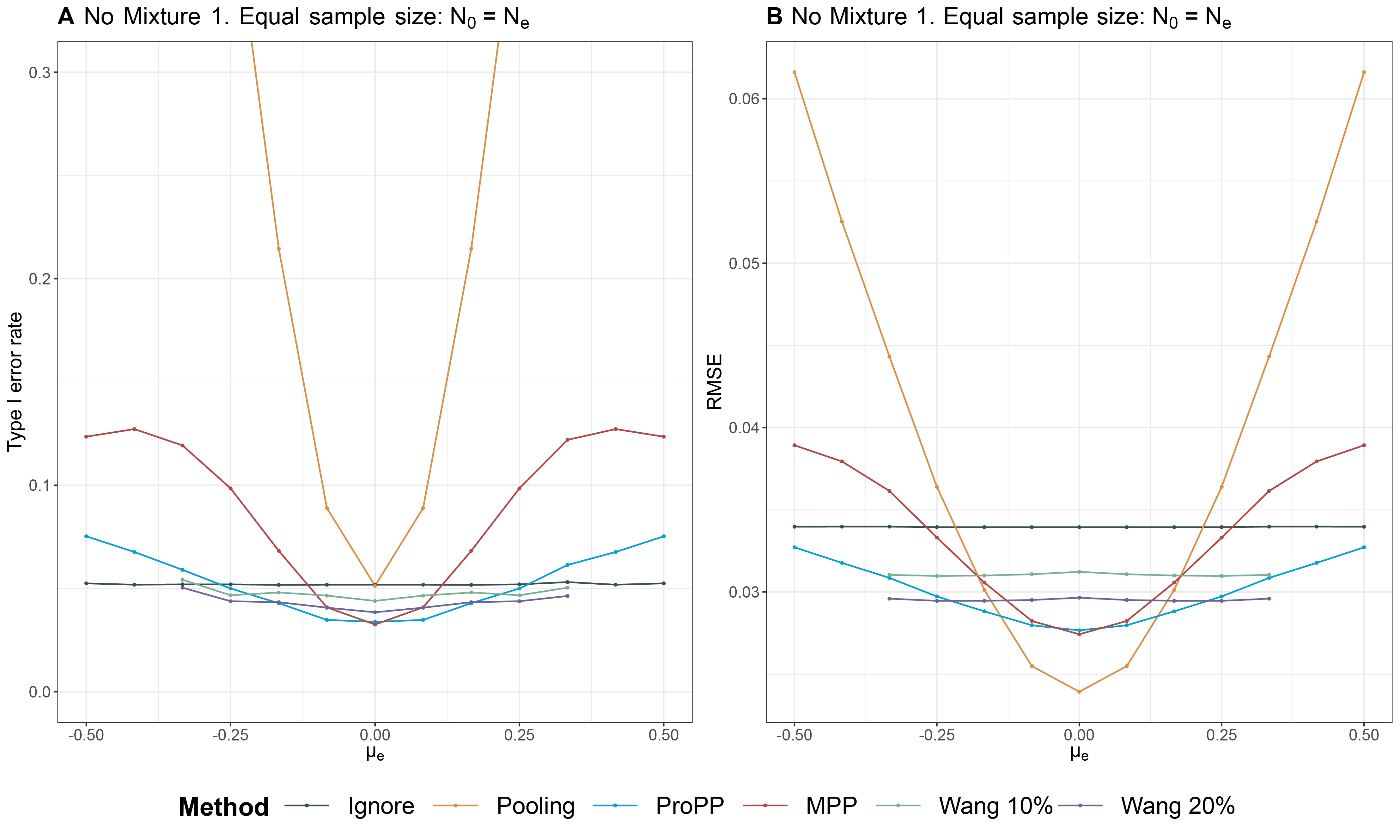}}
\centering 
\caption{Comparison of different estimation methods in terms of Type-I error rate (left) and root mean squared error (RMSE) (right) when there is no latent class structure in covariates (Setting 3). Estimates have been removed if $>5\%$ of the computations did not produce estimates.}
\label{Figure: No Mixture}
\end{figure}

We first explored how the methods would compare when there is no latent class structure in the distribution of the covariates, in the setting 'no mixture'. The results of this sensitivity analysis are depicted in Figure \ref{Figure: No Mixture}. Compared with the mixture setting (Setting 2), we observe a steeper increase in both RMSE and type I error rate due to the absence of the leveling effect caused by the latent class structure. Furthermore, the further the covariate distribution shifts, the less their overlap becomes. In case of a large difference in covariate distributions, the corresponding decrease in the number of similar patients rendered the algorithm of Wang et al.\cite{Wang2019b} unable to complete the simulations in a considerable number of cases; at the extreme $\mu_e = -0.5$ these methods did not generate output in 65\% of all simulations. We removed the line from the figures when the algorithm error rate exceeded 5\%. This result highlights a small advantage of weighting-based schemes over stratification-based approaches. In the ProPP, the weights are simply set to 0 when external patients have very different covariate values than the patients in the trial, implicitly discarding part of the data but allowing the analysis to continue. The rest of Figure \ref{Figure: No Mixture} shows increased RMSE and type I error inflation compared with \ref{Figure: Mixture}. The ProPP performs favorably compared with the MPP and Wang's suggested methods for a mild discrepancy e.g. $\mu_e \in (-0.25; 025)$ between the covariate distributions. All in all, in the absence of latent classes, the ProPP (i) fares reasonably well for small differences between covariates and (ii) accounts for larger distortions when covariate distributions overlap decreasingly. 

\subsubsection{Sensitivity analysis: superfluous covariates} \label{sec: superfluous covariates}

In Figure \ref{Figure: Superfluous Covariates} in the Appendix, we further examine the effect of including superfluous covariates, i.e. covariates that do not influence the outcome but do influence the allocation. We do this by setting $\beta_j = 0$, for $j = \{1\}, j = \{1,2\}, j = \{1,2,3\}, j = \{1,2,3,4\}$, whilst the overall effect of $\beta$ remains constant $(\sum \beta_{j} = c)$. 

We observe that the RMSE is relatively similar or merely increases slightly along with the number of 'superfluous' covariates included in our model. Without superfluous covariates, the lowest RMSE of the ProPP is 0.02742 and attained when $\mu_{\text{E}} = 0$ (see Section \ref{sec: Mixture}). The differences are almost negligible: when one covariate is superfluous, the RMSE increases to 0.02745 (0.1\%) and when three covariates are redundant, the RMSE increases to 0.02758 (0.6\%). The ProPP method seems to outperform the methods of Wang across the range of our simulation set-up when including redundant covariates, which suggests that the ProPP is relatively robust to misspecification. 

\section{Illustration: expanded access of vemurafenib for melanoma}
To illustrate our method in practice, we here jointly analyze data from the vemurafenib clinical trial and the vemurafenib expanded access program. Vemurafenib is a drug currently approved for the treatment of late-stage melanoma harboring a V600E BRAF mutation. The United States (US) Food and Drug Administration (FDA) approved vemurafenib in 2011 for patients who progressed on chemotherapy based in part on a single-arm Phase-II study ($N=132$).\cite{Sosman2012SurvivalVemurafenib} The European Medicines Agency (EMA) approved vemurafenib in 2012.

In addition to the regulatory studies, expanded access programs were set-up to grant patients unable to partake in the trials the opportunity to access vemurafenib prior to regulatory approval.\cite{Flaherty2014AStates} At 29 sites across the US, 371 patients received vemurafenib while simultaneously generating data on the treatment patterns, safety, and efficacy of vemurafenib in a real world setting. We obtained individual patient data from the trial and expanded access program through the data sharing platform Vivli. The data access request and analysis plan can be obtained online through Vivli.\footnote{https://vivli.org/combining-data-from-expanded-access-programs-and-conventional-clinical-trials-a-statistical-application-to-vemurafenib/}

The inclusion criteria of the expanded access program were less stringent than the criteria of the clinical trial, recruiting a broader patient population compared with the trial. For example, patients could only be included in the trial if they had an Eastern Cooperative Oncology Group (ECOG) performance score - a measure of physical fitness - of 0 or 1, whereas in the expanded access program, 19\% of the patients had a worse performance score of 2 or 3. Similarly, 75\% of patients in the EAP had stage M1c disease, meaning the cancer had spread throughout the body, compared with only 61\% of patients in the trial. Table 2 displays the differences in patient characteristics among a subset of (prognostic) variables across the trial and the expanded access program. For the expanded access program, 64\% ($241/371$) of expanded access patients had efficacy assessments available and were included in the analysis.

\clearpage 
\newcommand\MyIndent[1]{\hspace{1cm}#1}

\captionsetup[table]{labelformat=empty,skip=1pt}
\begin{longtable}{lcc} 
\caption{Table 2 Characteristics of the patients participating in the vemurafenib expanded access program (EAP) and trial.}  \\
\toprule 
 & \multicolumn{2}{c}{Clinical Program} \\ 
 \cmidrule(lr){2-3}
\textbf{Characteristic} & \textbf{EAP}, N = 241\textsuperscript{1} & \textbf{TRIAL}, N = 132\textsuperscript{1} \\ 
\midrule
Age at enrolment &  53 (13) &  50 (15) \\ 
Gender assigned at birth &  &  \\ 
\MyIndent Female & 95 (39\%) & 51 (39\%) \\ 
\MyIndent Male & 146 (61\%) & 81 (61\%) \\ 
Melanoma stage &  &  \\ 
\MyIndent M1a & 22 (9.1\%) & 33 (25\%) \\ 
\MyIndent M1b & 26 (11\%) & 18 (14\%) \\ 
\MyIndent M1c & 182 (76\%) & 80 (61\%) \\ 
Unresectable Stage III & 11 (4.6\%) & 0 (0\%) \\ 
ECOG performance status &  &  \\ 
\MyIndent Grade 0 & 112 (46\%) & 61 (46\%) \\ 
\MyIndent Grade 1 & 98 (41\%) & 71 (54\%) \\ 
\MyIndent Grade 2 & 30 (12\%) & 0 (0\%) \\ 
\MyIndent Grade 3 & 1 (0.4\%) & 0 (0\%) \\ 
Objective Response Rate & 129 (54\%) & 75 (57\%) \\ 
\bottomrule
\multicolumn{3}{l}{%
    \begin{minipage}{\linewidth}%
    \textsuperscript{1}Mean (SD); n (\%)%
    \end{minipage}%
}\\
\end{longtable}
\vspace{-5mm}

\subsection{Analysis and outcome}
\begin{figure}[ht!]
\centerline{\includegraphics[scale=0.5]{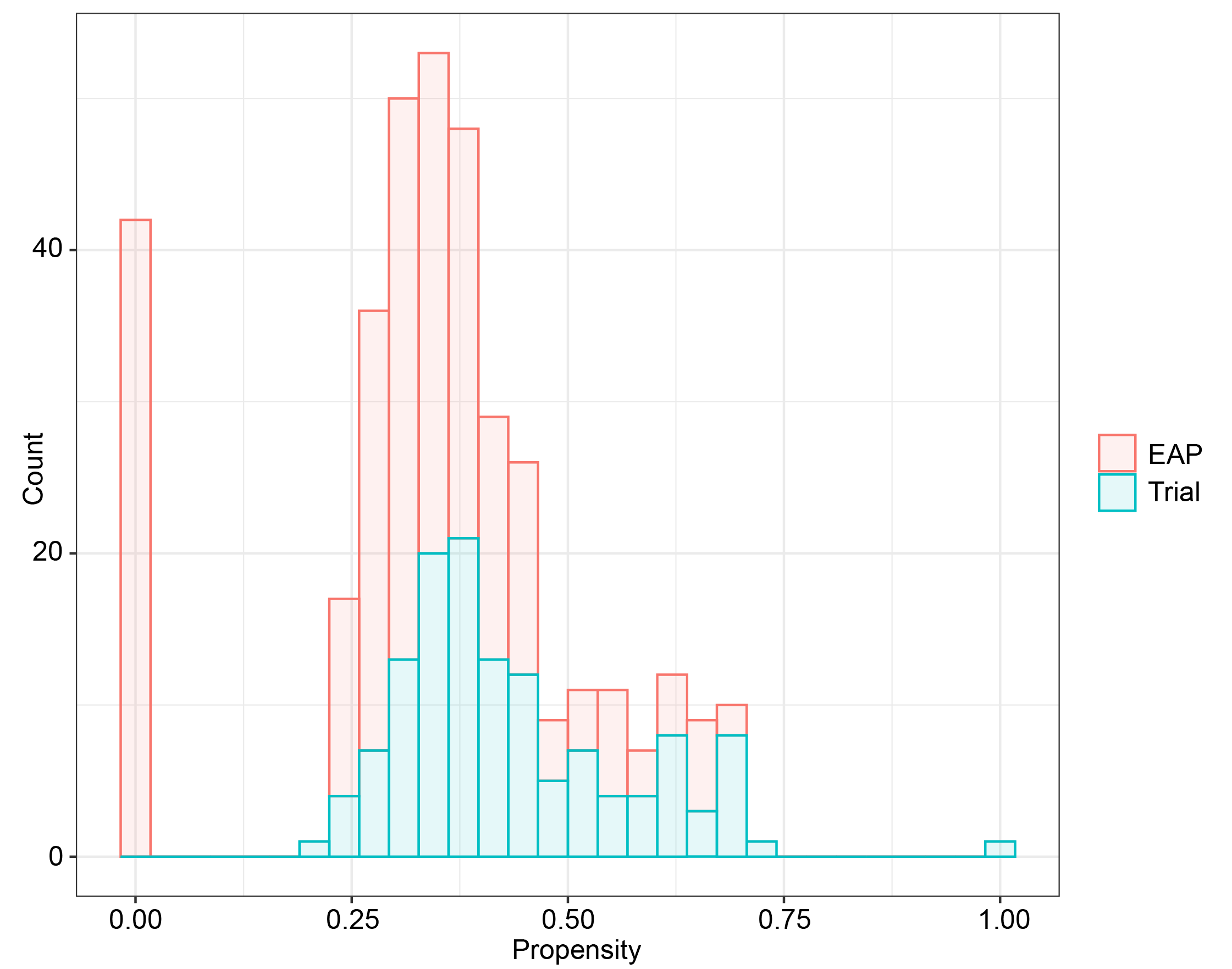}} 
\centering 
\caption{Distribution of propensity scores of patients in the vemurafenib trial and expanded access program (EAP).} \label{Figure: Propensities}
\end{figure}

In the first stage of our analysis, we estimate the probability of patients being in the trial conditional on their baseline characteristics. Frail patients (with a ECOG score $\ge 2$) were not allowed to participate in the trial and we expect these patients not to be integrated in our analysis. The propensities are depicted in Figure \ref{Figure: Propensities}. The 42 patients with weight 0 on the left hand side are indeed all 31 patients with ECOG 2 and 3, as well as 11 additional patients with a baseline melanoma stage of 'Unresectable Stage III' - the latter category was also not present in the trial. 

\begin{figure}[H]
\centerline{\includegraphics[scale=0.5]{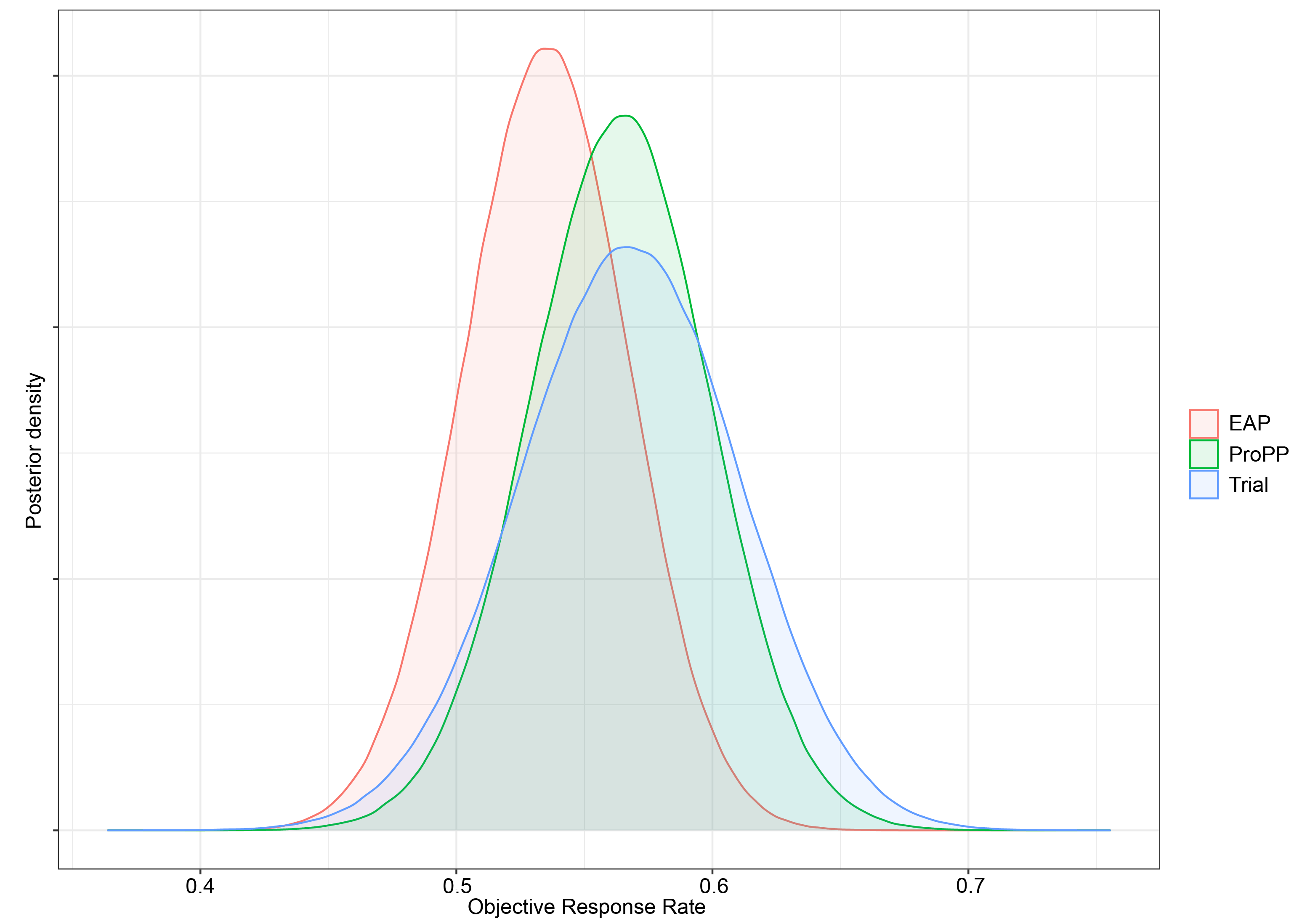}}
\centering 
\caption{Posterior distributions of estimates of the Objective Response Rate computed using data from the trial only, expanded access program (EAP) only, or combined EAP and trial through the ProPP method.} 
\label{Figure: Posterior EAPTRIAL} 
\end{figure}

The primary outcome of the trial and the expanded access program was the Objective Response Rate (ORR), defined uniformly as the fraction of patients with a complete response (CR) or partial response (PR). The estimate of ORR in the trial was 53\% (95\% confidence interval (CI): 44\%-62\%). In a Bayesian reanalysis, given a $U(0,1)$ prior for the ORR, the posterior is Beta(75+1, 132-75+1) distributed with posterior mean 56.7 \% and 95\% posterior credible interval (48.3\%, 65.0\%). The estimated ORR in the expanded access program was 54\% (95\% CI: 47\%-60\%), in a Bayesian reanalysis leading to a posterior that is Beta(129+1, 241-129+1) with a mean of 53.5\% and 95\% posterior credible interval (47.2\%, 59.7\%). The analysis with the ProPP leads to a posterior mean of 56.4\%, with a 95\% posterior credible interval of (49.4\% , 63.3\% ). The different methods are depicted in Figure \ref{Figure: Posterior EAPTRIAL}. 
We observe that the inclusion of expanded access data to augment the active trial arm leads to an increase in precision with a similar mean parameter estimate in this specific example of vemurafenib.
\newpage
\section{Discussion}

We developed a method to integrate the propensity score with variable power prior methodology. Our motivation stems from the increasing interest to incorporate external real-world evidence, and in particular expanded access data, into current trial data. Our novel ProPP method flexibly accounts for differences in outcomes and covariates between these two data sources in a two-stage design. Differences due to observed covariates are first incorporated through the propensity score. Remaining confounding is subsequently attenuated via the MPP in a dynamic borrowing setting. To our knowledge, we are the first to present the causal implications of the propensity score-integrated methods, and our modeling choices are guided by this causal interpretation. Our work explores the idea of augmenting treatment arms with current expanded access data. Overall, we observed that our method performs better than or on par with existing methods in a simulation study. 

In simulation our method provides higher precision (lower RMSE) compared with both 'naive' methods and 'hybrid' methods, at the cost of light-to-moderate inflation of type I error rate. The additional two-stage safeguarding does not lead to a significant loss when there is no difference in outcomes due to underlying differences in covariates. This finding is in line with previous research exploring hybrid two-stage designs.\cite{Liu2021, Han2017} Additionally, our method can be shown to behave similarly to the standard MPP when covariates are equal across data sets and, unlike the MPP, naturally accounts for differences in sample sizes between data sets. Compared with previous methods,\cite{Wang2019b, Chen2020,Lin2019} our method needs no pre-elicitation of a fixed power parameter or a fixed amount of external patients to be borrowed, nor does it require decisions on trimming, distance measures, or the number of strata. On the other hand, it does entail a choice of prior specification. This degree of flexibility of the ProPP leads to an increase in precision, but it comes at the cost of lacking an 'outcome-free' design principle as the Bayesian estimation of the power prior takes into account the posterior probabilities, whereas fixed power prior weights do not. When contrasts in outcomes are in part caused by contrasts in covariates, all propensity score-integrated methods outperform 'naive' methods - a conclusion backed by a recent review.\cite{Wang2022} Nonetheless, we echo prior scholarship that borrowing information entails a trade-off between cost (potential incremental errors in decision-making and type I error rate inflation) and benefits (increased precision, decrease in patient burden).\cite{VanRosmalen2018, Viele2014, Galwey2017} The high unmet need innate to expanded access programs together with the abundance of innovative statistical designs may tip the scale in favor of borrowing.

Our method, like all propensity score integrated designs, may be particularly applicable in a setting when a part of the external patients are similar to patients in the current trial. We argue that expanded access programs harbor these characteristics as typically two types of patients are included. The first category are patients who are excluded from the trial due to their baseline condition, e.g. when they are too frail to participate in a trial, but are nonetheless granted access out of 'compassion'. For these patients, the inclusion/exclusion in the trial will probably be driven by a difference in expected outcomes. The second category consists of patients who would have been eligible for the trial but are 'unlucky', as a trial is already fully enrolled, or as trial sites are geographically out of reach. Although analysing data of patients of the first category may lead to insights into the generalizability of treatments, it simultaneously may decrease the precision of the estimate and increase the chance of erroneous decision-making. Including the second category of 'unlucky', trial-like patients may on the other hand increase precision. Hence, an expanded access program may actually resemble the latent class simulation set-up, and we have shown that our method is able to correctly discriminate between these two classes of patients. Expanded access runs in parallel to ongoing trials, and these data hence form a 'current' external data source. This distinguishes expanded access data from 'historical' or 'non-current' external sources and limits the potential bias due to time trends. Furthermore, other scholars have suggested to explicitly incorporate the 'unmet medical need' or patient burden in trial design specifications - for example by adjusting the controlled type-I error rate in diseases with extremely low survival rates (e.g. glioma). As expanded access by definition is only available for patients with a high unmet medical need,\cite{Darrow2015, Chapman2019} this additional flexibility could be explored through the use of innovative statistical designs.

The similarity between data sources should play a decisive role whether to integrate external and current data and if so, to what degree. The transparency in hybrid two-stage methods using propensity scores allows one to inspect the balance of covariates across data sets before proceeding with the analysis. As such, it provides a quantitative addition to the qualitative measures suggested by Pocock.\cite{Pocock1976} The availability of a causal interpretation of the estimates, combined with the additional safeguarding in hybrid methods, altogether provides a statistically rational argument to attempt to include expanded access patients into decision-making. 

The acceptability of evidence synthesized from expanded access data in regulatory decision-making remains a topic of debate as these data are used in a qualitative, supportive manner. \cite{Polak2020, Rozenberg2020a, Polak2022, Caplan2016} Nonetheless, various regulators have put forth guidance on the (statistical) incorporation on 'real-world evidence'.\cite{FDA2019, Dreyer2018} In addition to statistical arguments, there are also ethical considerations of incorporating expanded access data: ignoring expanded access data would imply treating patients with investigational medicine without reaping the benefits of additional insights on the safety and efficacy for future patients. Lastly, it denies participating patients the freedom to altruistically advance clinical research. One could therefore argue that more attention should be devoted to the development of statistical methods to analyze expanded access programs. Our method provides a quantitative toolbox to augment treatment arms with expanded access data in a cautious and prudent way.

\subsection*{Limitations and future research}
We acknowledge several limitations to our study. First, we have chosen a subset of the potentially available methods for integrating propensity score and dynamic borrowing, and we did not consider other relevant comparator methods such as direct covariate adjustment.\cite{Banbeta2022}
Second, evaluating our method in terms of inflation of type I error rate could be questioned. For the true frequentist requiring strict type I error rate control, we know that given the external data, gains are typically not possible.\cite{KoppSchneider2020} For the true Bayesian, operating characteristics such as type I error rate are less relevant. Furthermore, these methods are a combination of frequentist and Bayesian methodology, as the propensity scores are still estimated from a frequentist logistic regression model. A fully Bayesian design that integrates the estimation of the propensity score remains uncharted territory.\cite{Kaplan2012, Chen2015} Our restriction of the propensity score weights is a result of this mixed methodology. It should be noted that our limiting of the weights to a maximum of 1, while possibly desirable from the point of view of a regulator, will result in weights that will potentially not be able to capture all of the confounding effects of the variables in the propensity score model.
Third, we derived the results from our method in the binomial setting. The binary outcome leads to a closed form posterior which greatly simplifies sampling and shortens computation time. We have not explored other outcome types, but it should be feasible to extend our method to time-to-event or normally distributed outcomes. Our method showed favorable computational performance compared with the method from Wang and others as described in their psrwe package,\cite{Wang2019b} where the propensity-score stratification sometimes failed to produce estimates.
Finally, our simulation set-up including latent classes was inspired both by the original simulation set-up of Wang and others,\cite{Wang2019b} and by our analysis of patient populations in expanded access programs. The latent class setup may however favor hybrid methods in our simulation. The underlying assumptions and plausibility of such specifications should be tested prior to utilizing our suggested approach. 

\subsection*{Conclusion}
We developed a novel statistical design to augment the treatment arm of a current trial with external (expanded access) data. We illustrated our method through causal interpretation, simulation, and a real-life application to expanded access data. Our study shows that our proposed method compares favorably with traditional and novel methods in simulation in terms of RMSE and type-I error rate, and may be a useful addition to the growing field of propensity score integrated dynamic borrowing approaches. The potential decrease in trial size and associated patient burden, the high unmet medical need in expanded access programs, together with the precautionary statistical set-up may favor the inclusion of expanded access with current trial data. Nonetheless, the inclusion of evidence sources remains a trade-off between bias due to including non-trial data and increased precision due to increased sample size.

\subsubsection*{Acknowledgements}
This publication is based on research using data from data contributor Roche that has been made available through Vivli, Inc. Vivli has not contributed to or approved, and is not in any way responsible for, the contents of this publication.

\subsubsection*{Author contributions}
TBP and JvR conceived the idea for this manuscript. TBP and JvR performed the statistical analysis. JAL provided the causal interpretation. TBP, JvR, and JAL drafted the manuscript. CAU-dG critically revised the manuscript. All authors approved the final version of the manuscript.

\subsubsection*{Data and code sharing}
The data used in this publication can be requested through Vivli. The code for replicating this study are available on the GitHub of the first author.\footnote{https://GitHub.com/TobiasPolak} 

\subsubsection*{Statements and declarations}
CAU-dG has received unrestricted grants from Boehringer Ingelheim, Astellas, Celgene, Sanofi, Janssen-Cilag, Bayer, Amgen, Genzyme, Merck, Glycostem Therapeutics, Astra Zeneca, Roche and Merck. TBP works part-time for expanded access service provider myTomorrows, in which TBP holds stock and stock options ($< 0.01\%$). TBP is contractually free to publish, and the service provider is not involved in any of TBPs past or ongoing research, nor in this manuscript. JVR and JAL do not report any conflicts of interest.

\subsubsection*{Funding}
CAU-dG, JvR and TBP work on a Dutch government grant from HealthHolland. For this grant, they research legal, ethical, policy, and statistical issues of evidence generation in expanded access programs (EMCLSH20012). HealthHolland is a funding vehicle for the Dutch Ministry of Economic Affairs and Climate Policy that addresses the Dutch Life Sciences \& Health sector.

\bibliographystyle{plain}
\bibliography{bibliography}

\section{Appendix Section}
\begin{figure}[ht!]
\centerline{\includegraphics[scale=0.40]{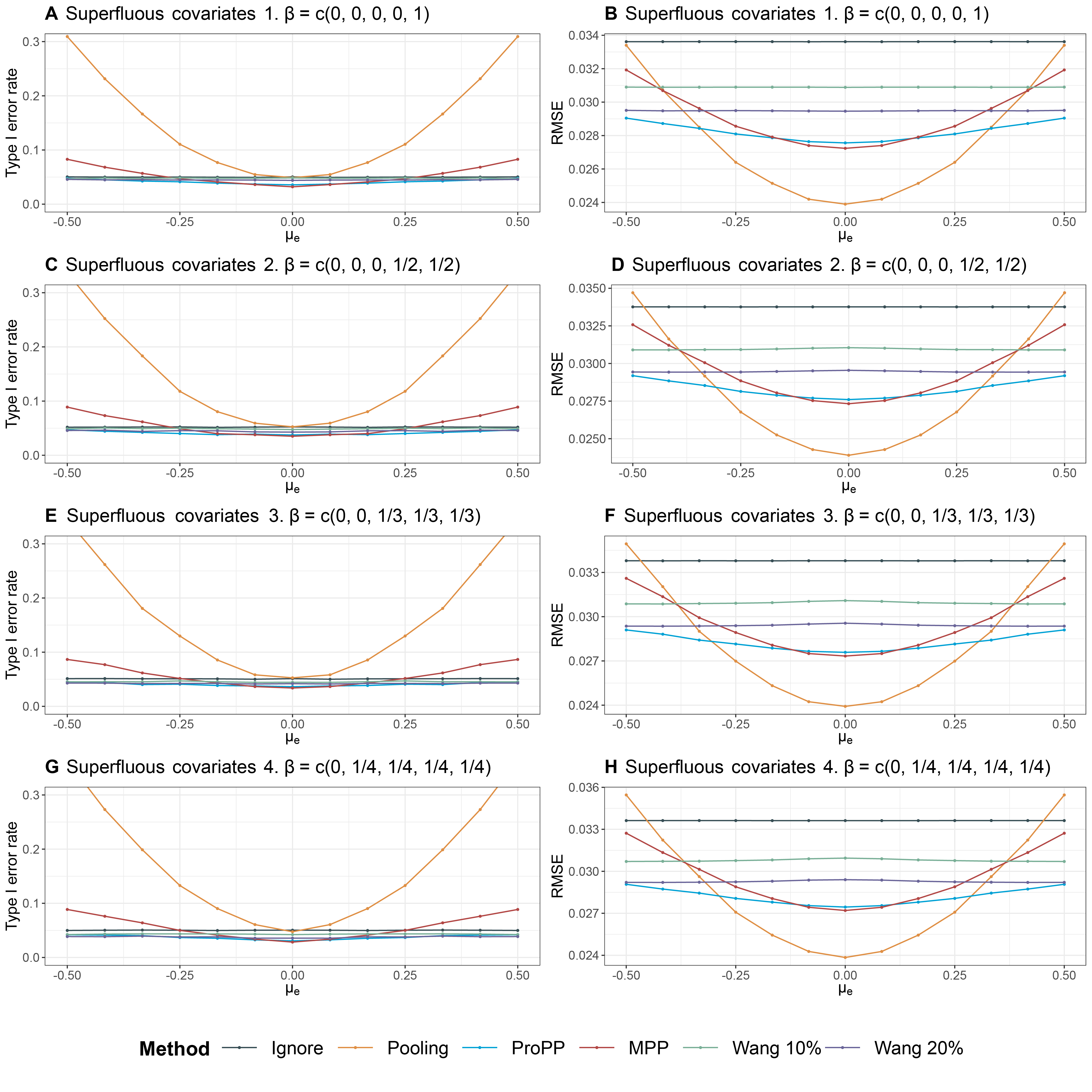}}
\centering 
\caption{Comparison of different estimation methods in terms of Type-I error rate (left) and root mean squared error (RMSE) (right) when the difference in outcomes is in part driven by difference in covariates. Some covariates do not influence the outcome $(\beta_j = 0)$ and should have been excluded. This is a sensitivity analysis (Setting 4).}
\label{Figure: Superfluous Covariates}
\end{figure}

\end{document}